# Assessing students' understanding of the concept of electric potential difference based on the SOLO taxonomy in upper- secondary students for a targeted assessment

Rahmani Fateme[1], Ahmadi Kalateh Fatemeh[2]
[1] Author: Faculty of Basic Sciences, Graduate Islamic Azad University, Tehran, Iran
[2] Correspondence author: Faculty of Basic Sciences, Associate Professor shahid Rajaee Teacher Training University, Tehran, Iran

## Abstract:

The concept of potential difference is fundamental to understanding electricity in upper-secondary physics courses. However, students often find it challenging to grasp. To address this, we designed and implemented a test aimed at assessing students' comprehension of potential difference. This researcher-developed test, with a reliability coefficient of $\alpha=0.74$, was administered to 92 students in grades 11 and 12. To evaluate the test and gauge the depth of student learning, we employed the Structure of Observed Learning Outcomes (SOLO) Taxonomy. Our study is the first to apply the SOLO Taxonomy to this physics topic, contributing new insights to educational literature. The analysis revealed that 71% of students displayed prestructural and unistructural levels of understanding—the shallowest learning stages—while only 3% achieved higher levels of comprehension. Additionally, our analysis of student responses identified areas where understanding diverged from scientific explanations. By analyzing these difficulties, educators can develop targeted assessment and implement similar practices to improve learning before these challenges become deeply ingrained.

Keywords: SOLO taxonomy, Physics Education, Electrical potential difference.

## Introduction

In upper-secondary physics education, students encounter many concepts related to electrostatics, which enable them to explain daily phenomena such as the operation of spray guns, inkjet printers, and air filters(Shen & Linn, 2011).

One of the most challenging concepts in physics is the electrical potential difference (PD). Teachers often observe challenges related to PD in science classes (Moynihan, 2018),(Taşkın & Yavaş, 2021). These difficulties can obstruct learning, and if not addressed at the elementary level, they may persist into higher education. While several studies have examined related topics like voltage in electrical circuit (Liu et al., 2022), further research on potential difference is necessary (Arslan, 2006; Maloney et al., 2001).

The challenges in this area are due to a lack of awareness of students' diverse backgrounds. Carl Bereiter's "three worlds" concept—physical, social, and symbolic—offers insight into the learning process(Bereiter, 2005). These interconnected worlds suggest that learning in one can enhance learning in another. It is crucial to recognize that students bring pre-existing realities to class, which can hinder their learning if not properly assessed beforehand (Jaber et al., 2023)Studies show that teachers ask hundreds of questions daily, most of which require simple factual recall (Panicker et al., 2019). For example, Gall found that 60% of teachers' questions required factual recall, 20% were procedural, and only 20% encouraged student thought(Moje, 2007). Students are more likely to engage in deep thinking when teachers use rich, higher-order questions and explore students' mind maps(Asadi et al., 2017); Consequently, addressing the challenged in this manner could reduce their prevalence (Adeniji et al., 2022).

In previous studies, SOLO taxonomy examined the level of deep learning among students in various fields, including chemistry(Tian et al., 2024), mathematics(Adeniji et al., 2022; Mulbar et al., 2017) . Given the similarities between teaching mathematics and physics, the SOLO taxonomy can also be effectively applied in physics education. In this study, we attempt to ask rich questions about electric potential difference using the SOLO taxonomy to evaluate students' learning depth. These tasks focus on evaluating the background of students and promoting creativity and tasks that are not necessarily difficult but create deep learning (Potter & Kustra, 2012).

The study began with an extensive review of relevant literature. We focused on research involving SOLO-based assessment in educational contexts and explored studies addressing challenging area in the understanding of potential difference. After establishing a clear framework, we defined the research objectives with precision. The subsequent section detailed the data analysis methods employed, ensuring a thorough explanation of our approach. Finally, we presented the results of our analysis, highlighting the key findings and their implications.

**Literature Review**

The purpose of this study is to evaluate the extent of students' understanding of electricity concepts. In this session, we will review previous research on the developing assessment tools, SOLO taxonomy and the concept of potential difference.

Evaluating the depth of learning among students is crucial for fostering higher-order cognitive skills and ensuring meaningful educational outcomes. Moore's study(Moore et al., 2024) emphasizes that assessment of learning outcomes should not merely evaluate rote recall but instead prioritize questions that challenge students to engage deeply with concepts. By designing assessments aligned with higher levels of the SOLO and Bloom taxonomy—such as the relational and extended abstract stages—educators can encourage students to connect ideas, analyze relationships, and synthesize knowledge into broader frameworks. Nieminen (Nieminen et al., 2024) also highlights the critical role varied forms of formative assessment in supporting student inquiry during science lessons.

SOLO taxonomy is a valuable framework for helping students achieve a deeper understanding of a subject. This taxonomy, introduced by Biggs and Collis in 1982(Biggs & Collis, 1980), stands for the Structure of Observed Learning Outcomes. Originally, they demonstrated its applicability in teaching history, mathematics, and geography. Since then,

SOLO has been employed to assess knowledge in various fields, including computer programming (Aronshtam et al., 2021; Malik et al., 2021), the C language(Shi et al., 2017) , solving mathematical problems (Silwana et al., 2021)and science education(Irvine, 2021). But The SOLO model is being used for the first time in a physics context with this study.

One important tool for assisting students in gaining a deeper comprehension of a subject is the SOLO taxonomy. When students can answer questions at a high (abstract) level, it demonstrates their ability to apply learned knowledge to new, unaddressed scenarios(Ledezma et al., 2023) .This approach encourages students to engage with challenging questions, preparing them to face real-life challenges (Davies & Mansour, 2022).

Numerous studies have explored the application of SOLO in education. For instance, research in mathematics analyzed students' reasoning abilities in solving statistical problems using the SOLO taxonomy. The results indicated that only high-achieving students could address tasks at an extended abstract level (Potter & Kustra, 2012).The SOLO taxonomy provides a systematic framework to describe student performance, especially in school tasks. Similar studies have shown that SOLO categorizes learners' cognitive skills effectively, aiding in understanding the complexity of problem-solving (Mulbar et al., 2017). Building on these findings, this study examines the applicability of SOLO in physics. Table 1 presents the five levels of SOLO taxonomy, illustrated by students' responses to a question on electric potential difference.

Table 1: Levels of SOLO taxonomy with a sample question from the concept of electric potential difference

| |
|---|
| Question 3-A: an electron with an initial speed of 200,000 m/s is brought to rest by an electric field. Did the electron move into a region of higher potential or lower potential? |
| Answer at the unistructural level: by the formula, it is **defined** that the potential difference is related to a change in the speed charge. Answers at this level contain only some memorized formulas. |
| Answer at multistructural level: to evaluate the potential difference, the **combination** of varied features should be considered: the velocity must be considered. The speed change is negative in this question as it decreases from 200,000 to zero. |
| Answer at the relational level: the speed of the electron is decreasing. So, **that is why** this motion requires work to be done. The electron moves into a negative surface (with a lower potential). |
| Answer at extended abstract level: since the electron's velocity is declining, the negative charge moves from the positive surface (with a higher potential) into the negative one (with a lower potential). Thus, the particle **is predicted** to move from a higher potential to a lower one. |

Table 1 shows the answers at different levels to question 3-A. The **emboldened terms** highlight the verbs employed within solo taxonomy. The scoring system for these answers is explained in Table 2 (page 10).

Figure 1- Solo taxonomy symbols and verbs, Produced by Pam Hook

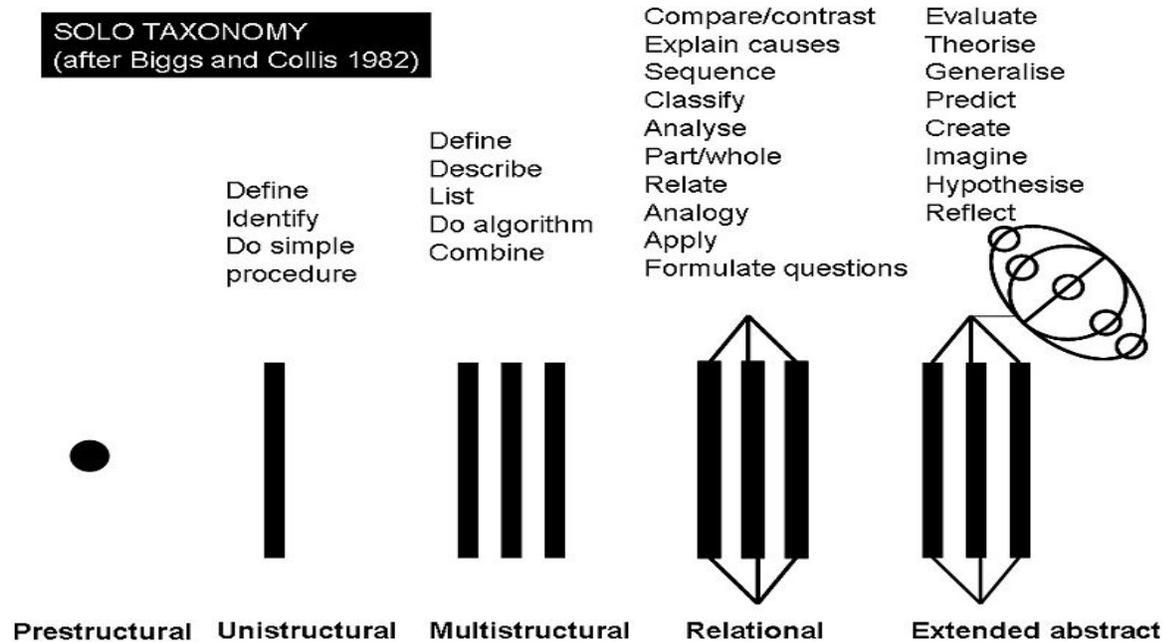

Figure 2- Multistructural tasks produced by Pam Hook

In field of science education , Rusek (Vojíř & Rusek, 2019) conducted a systematic review of science textbooks evaluation published from 2000 to 2018. Their analysis reveals that Biology and chemistry dominate, while only 8% of studies address physics, highlighting the need for more emphasis on evaluating physics textbooks.

Electrostatic forces and fields were also identified as challenging topics in science textbooks(Mazibe & Rollnick, 2024) . This study contributes to broader discussions on the adequacy of instructional materials in addressing students' misconceptions and fostering a deeper understanding of physics concepts.

Similarly, potential difference is a challenging concept for students to grasp, prompting extensive research in this area. A study using a 24-question multiple-choice instrument assessed students' understanding of electrostatic concepts. The findings revealed that students struggled to comprehend that changes in the path of a particle do not affect the electric potential difference between two points(Mi et al., 2023) . This concept was specifically evaluated in question 4-A of our study. Additionally, a study in Spain analyzed students' explanations at both macroscopic and microscopic levels in electrical circuits. The results suggest that traditional teaching methods are particularly ineffective at addressing microscopic phenomena, such as charge movement in an electric field (Leniz et al., 2017).

**Research Objects**

The study examines students' comprehension of potential difference using SOLO taxonomy to assess their understanding. Challenging areas about the concept were identified. By analyzing these difficulties, educators can develop targeted assessment. Focusing on areas with challenges allows teachers to optimize instruction time in sections where understanding is stronger. The research questions were:

1. To what extent can SOLO taxonomy-based instruction explore the depth of student learning?

2. What are the common challenges in the physics syllabus regarding potential difference?

**Research Method**

This section describes the research participants and outlines the procedures for conducting the exam, including both the pilot and formal test phases.

*Research design*

To initiate this study, we conducted a thorough review of the concepts of electric potential difference and electric potential energy as presented in high school textbooks for grades 11 and 12. We then determined the expected level of student comprehension based on Bloom's taxonomy, aiming to align our assessment methods with these educational standards.

*Participants*

The study was carried out in December of the academic year 2023–2024 in a public high school. Students typically learn about potential difference (PD) in the first semester of grade 11, and they revisit the topic in grade 12 in preparation for their final exams. Consequently, we selected this group of female students as our target group, as they had recently covered this topic in their classes. Mastery of this syllabus content significantly impacts students' overall academic performance in physics and their future studies.

Despite the relatively small sample size, the selection was informed by the combined expertise of the research team's teachers and insights from previous studies with similar sample sizes(Taşkın & Yavaş, 2021),(Davies & Mansour, 2022),(Elassabi & Kaçar, 2020) .The prior studies involved 18 and 30 students, respectively. Our study included 92 students from grades 11 and 12, aged 17 and 18.

*Design and development of test*

In the next step, we set questions at various levels, guided by the SOLO taxonomy.

*Initial item selection of questions:* We reviewed grade 11 textbooks and relevant practices with input from our expert group.

Figure 3: The proportion of questions in the bloom category

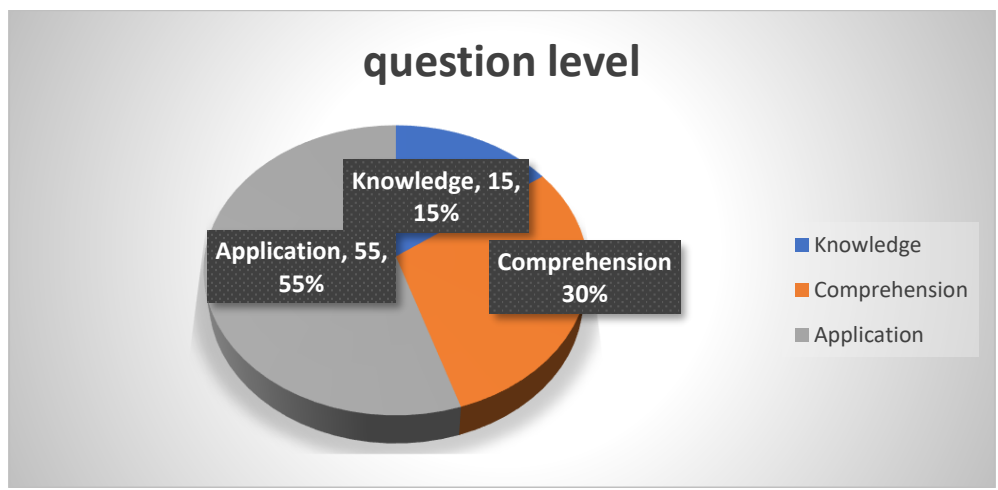

Figure 3 illustrates that more than 50% of the tasks in the high school textbook's section on electrical potential difference are categorized at the application level. Additionally, 15% of the tasks fall under the knowledge level, while 30% are comprehension questions. These distributions enabled us to formulate higher-level questions, including those at the extended abstract level, for our study.

We chose SOLO taxonomy over Bloom's taxonomy for assessing the depth of learning outcomes for several reasons:

1. While Bloom's taxonomy is a valuable tool for lesson planning, SOLO taxonomy is more effective for summative and formative assessments(De Florio, 2023),(Aripin et al., 2020).

2. Bloom's taxonomy suggests a fixed relationship between questions and students' responses (Jones et al., 2009). In contrast, SOLO taxonomy allows for different levels of questioning and answering.
3. Each answer in the SOLO taxonomy is assigned a specific score, simplifying the grading process for educators.
4. Bloom's taxonomy presents a sequence of complexity across its levels (Krathwohl, 2002), whereas SOLO taxonomy does not assume increasing difficulty.
5. Research comparing the reliability of the two taxonomies found that SOLO taxonomy is more reliable in classifying cognitive levels in assessment questions(İlhan & Gezer, 2017).
6. Whereas Bloom's taxonomy emphasizes a progression in cognitive difficulty, SOLO taxonomy focuses on the depth of understanding(Keane et al., 2016) .
7. Following the identification of limitations within Bloom's framework, SOLO taxonomy has been applied to diverse educational assessments(Page, 2022).

The test provides a framework for analyzing a student's depth of knowledge. To create five descriptive questions, we evaluated various textbooks and selected practices from established sources(Redish et al., 1998),(Knight, 2015),(Hewitt et al., 2003). These questions must not only align with students' abilities but also adhere to current educational standards. The questions are categorized at different levels of difficulty, as detailed in Table 1.

*An expert review:* This was conducted to assess the validity of the questions, involving a group of educators: two researchers and one teacher with an extensive background in physics education. The questions were improved and refined greatly by the knowledge and experience of this expert group.

Table 2: levels of five descriptive questions and scoring system

| question | level | mark |
|---|---|---|
| 1-A | unistructural | 1 |
| 1-B | multistructural | 2 |
| 1-C | multistructural | 2 |
| 1-D | unistructural | 1 |
| 2-A | abstract | 4 |
| 3-A | relational | 3 |
| 3-B | abstract | 4 |
| 4-A | relational | 3 |
| 4-B | abstract | 4 |
| 5 | multistructural | 2 |

The first column in Table 2 categorizes the two questions (1-A, 1-D) as unistructural. This is followed by three multistructural questions. Additionally, two questions are classified as relational, and three are at the abstract level. Question 1 comprises four parts (A-D), each addressing different levels from unistructural to multistructural. Questions 3 and 4 consist of two parts each. The third column of Table 2 provides the scoring system: unistructural answers receive a score of 1, while abstract answers are awarded a score of 4.

*Pilot test*

The test comprised five descriptive questions. To evaluate the validity and efficiency of this exam format, we conducted a pilot study involving thirty students: twenty from grade 11 and ten from grade 12. The test incorporated various learning levels, as detailed in Table 2.

To make sure the test worked, a content validity analysis was done. A Cronbach's alpha coefficient of 0.74 was found after statistical analysis of the data, indicating strong internal consistency and good reliability(Tavakol & Dennick, 2011). The test's ability to evaluate students' comprehension at various SOLO levels was effectively verified by the results of the pilot test.

*Formal test*

Students were given the formal exam, and they were told to submit thorough, detailed answers. To ensure accuracy, students were informed that their scores would positively impact their formative assessment results.

The study involved 62 additional students, with 32 in grade 11 and 30 in grade 12. The marking process was carried out by three teachers following these steps:

1. Teachers independently graded the papers using a pre-defined answer key, as outlined in Table 2. This key, based on the criteria in Table 1, guided the scoring for each question.
2. After grading, the mean score was calculated by reviewing the teachers' evaluations. This two-step approach ensures a more consistent and fair grading process (Rashid & Mahmood, 2020).

Figure 4. A summary of the methodology

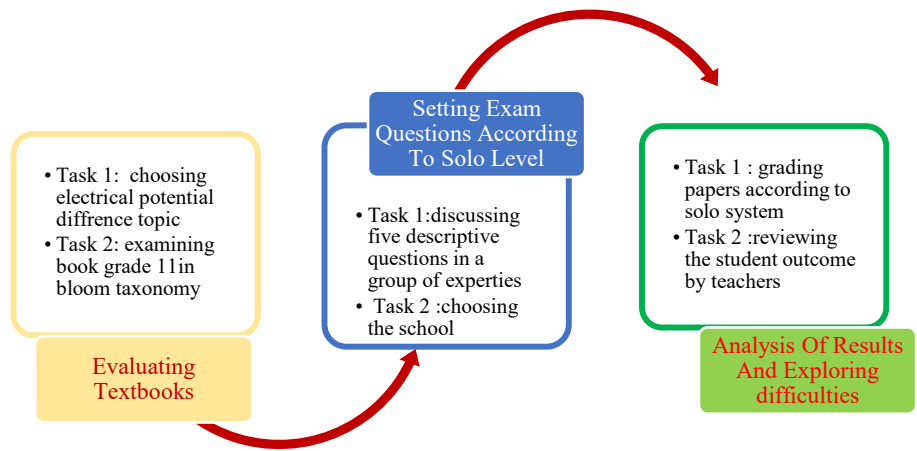

**Date Analysis**

In a public high school with 92 female students, the study was carried out. Two stages of the data analysis were implemented. In the first phase, a detailed, question-by-question analysis was performed on students' responses to five descriptive questions. In the second phase, a comprehensive analysis of the responses was undertaken to identify trends across the entire dataset.

*Analysis of participants' responses*

The evaluation exam was conducted during a session collaboratively managed by two teachers and a researcher. Participation in the exam was entirely voluntary. The scoring of students' responses allowed for the determination of specific scores for each question and overall scores. This process was conducted entirely anonymously.

A grading system, detailed in Table 2, was used to evaluate the answers. Each response was assigned a score based on its level: a unistructural response received one point, while a multistructural response received two points. This analytical approach aimed to identify disparities or trends in performance at different levels, enhancing our understanding of students' learning trajectories.

*1)The initial analysis of data*. This step allowed us to closely examine the distribution of responses for each question.

Question 1: Charge $q=-4\mu$ is moving with constant speed in an electrical field of $5 \times 10^5 Nc^{-1}$.

Figure 5: question number 1

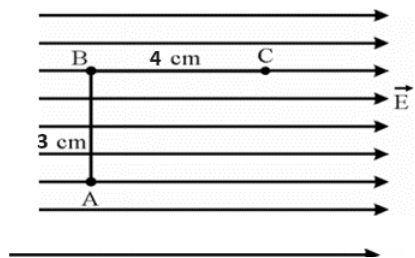

1-A: Calculate the force amount.

1-B: Calculate the work done in the AB path.

1- C: Calculate the work done in the BC path.

1-D: What is the whole potential energy difference in this displacement?

Questions 1-A and 1-D are unistructural because they require only memorized facts to be retrieved. Question1-B and 1-C are at the multistructural level because students need to identify the steps involved to do the task.

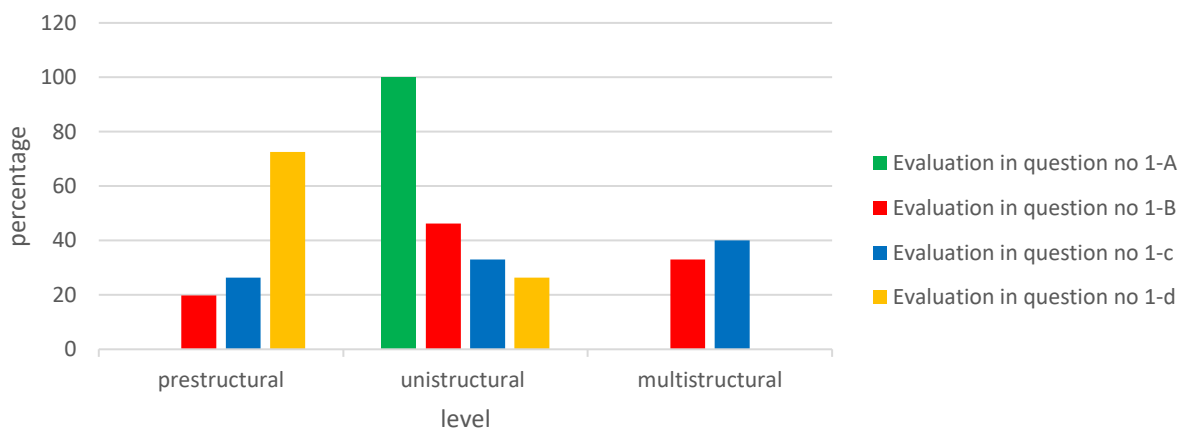

Figure 6: Evaluation in question number 1

Figure 6 illustrates that in question 1, most students struggled with parts D, C, and B, while part A was answered correctly by all students. For parts B and C, around 40% of responses were correct, indicating these questions were at the multistructural level. Conversely, approximately 20% of the answers to these parts were incorrect, reflecting a pre-structural understanding. In part D, around 70% of students were unable to respond correctly, also indicating a prestructural level

Question 2: A proton is released from rest at the positive plate of a parallel plate capacitor. It crosses the capacitor and reaches the negative plate at a speed of 50,000 m/s. The experiment is repeated with a He+ ion (charge $e$, mass 4 $u$). What is the ion's speed at the negative plate?

This question is situated at the abstract level since students are required to evaluate varied formulas containing mass and speed. In so doing, they can predict the correct responses to the task.

Question 3-A. An electron with an initial speed of 200,000 $ms^{-1}$ is brought to rest by an electric field. Did the electron move into a region of higher potential or lower potential?

Question 3-B. What was the PD that stopped the electron?

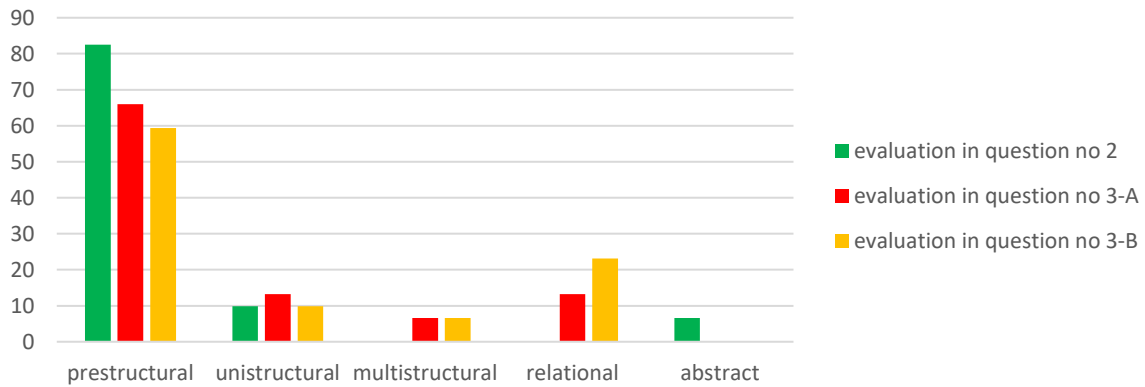

Figure 7: Evaluation in question number 2, 3

Figure 7 illustrates student performance on three exam questions (2, 3-A, and 3-B), which varied from prestructural to abstract levels. Question 2, identified as one of the most challenging due to its abstract level, was answered incorrectly by 82.5% of students, indicating a prestructural understanding. Similarly, in questions 3-A and 3-B, approximately 60% of responses were incorrect, also reflecting a prestructural level of comprehension. These results suggest that students consistently struggled with these concepts.

Question 3-A was explained in detail in Table 1 and question 3-B is at a relational level because students need to show their understanding by applying formulas which include recognizing the relation between variables and integrating them in one formula.

Question 4-A: two protons are launched with the same speed from point 1 inside the parallel-plate capacitor. Points 2 and 3 are at the same distance from the negative plate. Is $\Delta U$, the change in potential energy along the path 1 - 2, larger than, smaller than, or equal to the path 1-3?

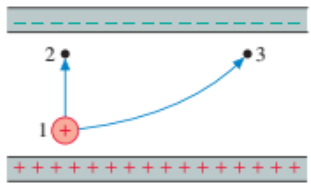

Figure 8: question number 4

The process of solving this problem involves assessing various formulas containing distance and potential energy to determine the correct one and create the one that theoretically explains the codependency of potential energy from the path taken.

Question 4-B: Is the proton's speed $V_2$ at point 2 larger than, smaller than, or equal to $V_3$? Explain.

This question falls within the relational level of solo taxonomy because the students need to explain the relation between distance and the fact that they have the same speed.

Question 5: Rank in order, from largest to smallest, the electric potentials $V$ (a) to $V$ (e) at points a – e.

This question is at a multistructural level since students should make a list of points and order them based on the distance to the surface.

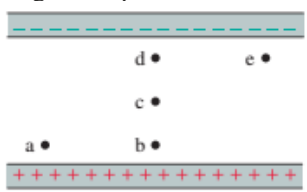

Figure 9: question number 5

Figure 12: Evaluation in question

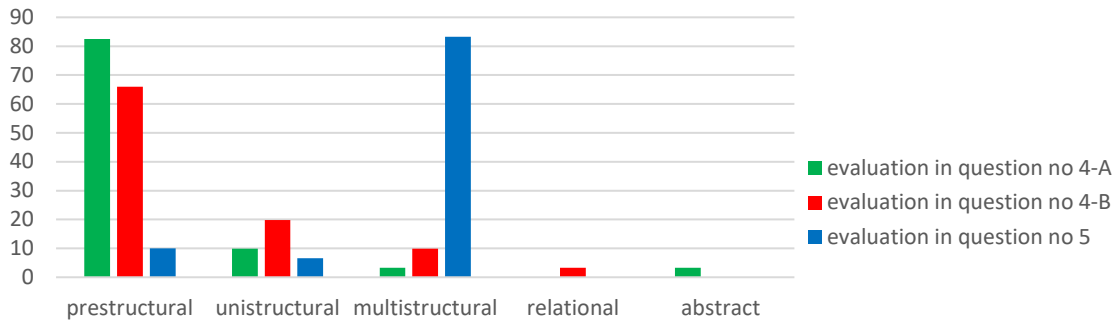

Figure 10 illustrates student performance on three questions (4-A, 4-B, and 5). Question 4-A proved challenging, with 82.5% of students scoring at the prestructural level, indicating incorrect responses. Similarly, Question 4-B, which was set at an abstract level, saw only 3.3% of students answering correctly, revealing challenging area. In contrast, Question 5, which was presented in a familiar context, was correctly answered by 83.3% of students, reflecting a multistructural level of understanding. These results suggest that students demonstrated a low level of comprehension on Questions 4-A and 4-B.

2)Students' interview:

To gather more information about the causes and extent of challenging area of understanding, we conducted interviews with two students. We asked each student three questions: the first two were identical for both, while the third differed. During the first interview question, which explored students' prior knowledge and theoretical concepts, neither student showed significant difficulty. Concepts discussed included uniform fields, formulas related to potential difference (PD) changes, potential energy, and work done. In the second question, both students demonstrated a clear understanding of the scenario, and the questions posed.

In question 2 of the test: In the third question of the interview, which was related to question 2 of the tests, we asked about the formula relating kinetic energy to changes in potential difference and how force and distance are incorporated. Although reciting the formula appeared straightforward for Student 1, she struggled with the mathematical application.

The purpose of this question was to evaluate the students' understanding of predicting particle motion in an electric field, considering the relationship between electrical potential difference and kinetic energy. Students were expected to know formulas and calculations for converting potential energy into kinetic energy. This knowledge allows them to determine the speed of particles with varying masses under consistent conditions. Student 1 successfully wrote all relevant formulas.

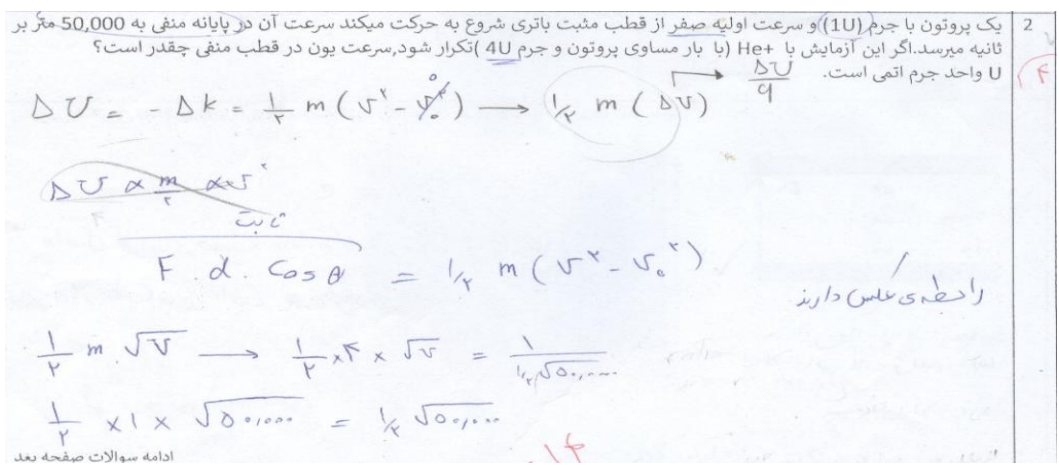

Figure 11: Here is the student 1 answer sheet. The student demonstrated an understanding of inverse proportionality between mass and speed (based on the formula of kinetic energy). However, she appeared to encounter confusion using mathematical concepts related to ratios and square roots.

Mathematical errors made by students can be categorized into several types: calculation errors, procedural errors, and symbolic errors. For instance, consider a problem involving the speed of different charges moving across an electric field.

$$\Delta K = -\Delta u$$

$$F.d.\cos\theta = {1}/{2}\, mv^2$$

The change in speed of an electric particle is dependent on its mass. Specifically, if the mass of the particle is quadrupled, its speed will be halved. Although Student 1 recognizes this as a basic mathematical principle she has learned, she struggles with visualizing particle scenarios and applying this concept in physics (Redish & Kuo, 2015). Additionally, we interviewed Student 2 and analyzed his response to question 4-A.

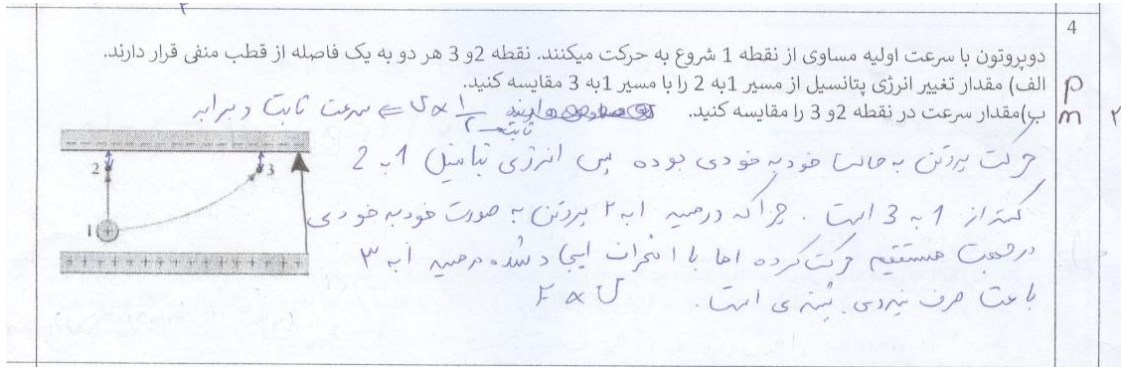

Figure 12 presents a student's response to question 4-A. The potential difference (PD) between two points in an electric field is determined solely by the vertical distance between them. The electric field is constant along the charge's path, so it makes no difference which direction the charge takes to complete the task. Students often struggle with this concept because they tend to focus on the movement of the charge itself, rather than the underlying principles governing the field(Mi et al., 2023) .

In Question 4-A, the objective is to evaluate the learner's understanding of the relationship between the starting and ending points in a uniform electric field and the associated changes in potential energy, irrespective of the path taken. Student 2 responded that a longer path requires more force to overcome, leading to a larger change in potential difference. While this answer partially addresses the changes in potential difference, the student appears confused about the role of force in these changes.

Interviews revealed that students need additional tools to visualize abstract concepts, such as potential difference, to better overcome their difficulties. Solutions such as Augmented Reality (AR)(Ropawandi et al., 2022) and design constraints(Magana et al., 2021) have been suggested by other studies to address these challenges.

*Total analysis of outcomes*

An analysis of student performance revealed that the lowest exam score was 3 out of 20. A significant majority of students (72%) scored between 2 and 6 points. This distribution suggests a notable clustering of scores towards the lower end of the scale, indicating potential challenges in understanding the material.

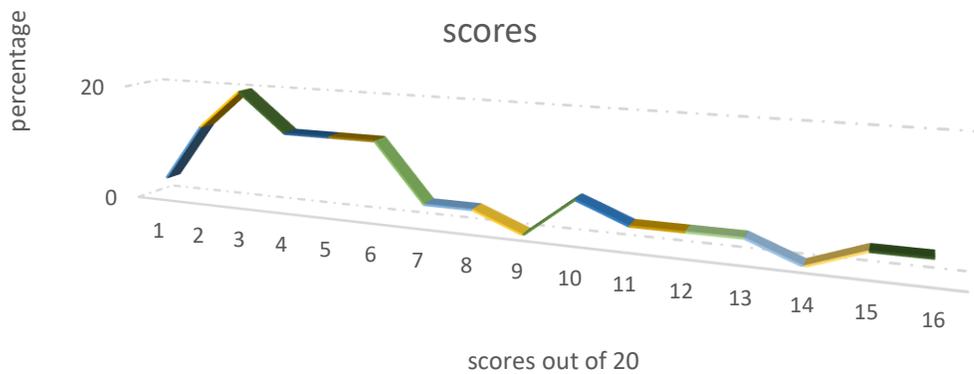

Figure 13: The students' outcomes in the test

The distribution of students' scores is illustrated in Figure 13. The horizontal axis (x-axis) displays the scores students received, ranging from 1 to 16, while the vertical axis shows the percentage of students who achieved each score. Most students received a score of 3, which was obtained by 20% of the students. Scores of 2, 4, 5, and 6 were each achieved by 13.2% of students. In contrast, only 3.3% of students attained the highest score of 16. This distribution indicates that most students struggled to reach higher levels of learning in the classes.

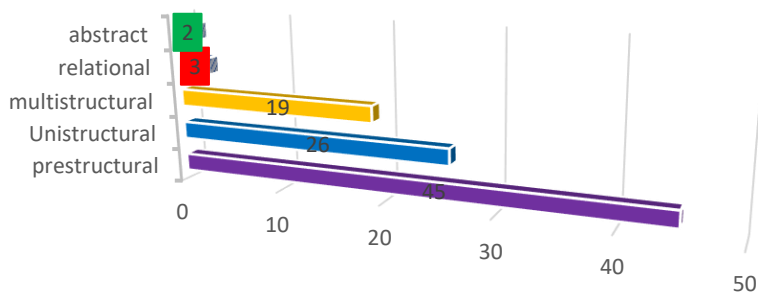

Figure 14: Percentage for every level

Figure 14 illustrates the distribution of student responses according to the SOLO taxonomy. A significant portion of the responses fell into the prestructural level, indicating that many students did not fully understand the topic. In contrast, only 2% of students achieved the abstract level, demonstrating a deep comprehension of the material. This distribution suggests a need for instructional strategies that better support students in reaching higher levels of understanding.

The results show that many students struggle to achieve a deep understanding of the material in this syllabus of schoolbooks. The finding aligns with other studies that SOLO taxonomy can be a useful taxonomy to assess learning outcomes(Davies & Mansour, 2022; Irvine, 2021).

**Results and discussion**

The study applied SOLO Taxonomy to evaluate students' understanding of electrical potential difference at the senior high school level. Tests were administered to 92 students in grades 11 and 12, with a reliability score of α=0.74. The results revealed that 71% of the students lacked a deep understanding of the concept. The study highlighted specific areas where students struggled, offering insights for educators to enhance their instructional approaches and promote deeper learning.

In response to the research question 1—how well does SOLO Taxonomy-based instruction explore the depth of student learning? The findings indicate that many students struggle with the material. Only 3% of the students achieved high learning levels, with most remaining at the prestructural and unistructural levels, the lowest on the learning scale. These results suggest that significant effort is needed to improve understanding in this area of physics. These findings align with other studies, as shown in Table 3, indicating that the SOLO Taxonomy effectively explores the depth of learning on this topic.

Table 3: challenging concepts about electric potential difference

| Challenging concepts | Question | References that confirm the item |
|---|---|---|
| 1) The speed change of a charge across the electric field depends on the path formed between two surfaces. | 4-A | Mi, huaishuai, et.al(Mi et al., 2023) Maloney, David. et al(Maloney et al., 2001) |
| 2) Students usually have difficulties understanding how the electric potential of a charge changes when it moves in an electric field. | 3-A | Mi, huaishuai, et.al(Mi et al., 2023) |
| 3) the work performed by electric force and the way that motion of a charge in an electric field changes electric potential energy. | 4-A | Burde & Wilhelm (Burde & Wilhelm, 2020) |

To address Research Question 2: What are the difficulties of students in this physics syllabus? The analysis reveals:

1. **Common Errors in Question 4-A.** Students frequently make mistakes in this question. Textbooks clarify that the angle formed by displacement and force determines the amount of work completed. However, 82.5% of students (Figure 12 in previous pages) could not provide the correct response, confusing the concepts of path and angle. This confusion led to misunderstandings about how changes in the path do not affect the final speed at the destination.

2. **Challenging area about Charge Movement.** Any charged particle, regardless of the kind of charge, accelerates as it moves in the direction of the potential field, gaining kinetic energy and losing potential energy. This process is analogous to a rock falling under gravity, where gravitational potential energy decreases as kinetic energy increases. Despite this, 66% of participants (Figure 10 in previous pages) struggled to analyze the motion of particles in an electric field, leading to incorrect answers. These difficulties in visualizing the motion of a charge and understanding the relationship between potential difference and velocity were also observed in a study involving 1,850 Chinese high school students(Mi et al., 2023). Another study with 5,000 students in introductory physics classes confirmed similar difficulties(Leniz et al., 2017) .

The findings of this study are consistent with prior research, particularly Moore's conclusion(Moore et al., 2024) that assessments targeting higher-order thinking skills effectively measure the impact of diverse instructional methods. Our analysis, using the SOLO taxonomy, highlighted significant gaps in students' conceptual understanding of potential difference. Moore's findings suggest that structured interventions and assessments emphasizing application and analysis could help bridge these gaps. Prioritizing assessments that reward critical thinking over rote memorization enables educators to support students in overcoming conceptual challenges and achieving deeper comprehension.

This study also aligns with Nieminen's work(Nieminen et al., 2024) on the value of real-time formative assessment in addressing conceptual misconceptions. While our research focused on structured assessments, Nieminen highlights the complementary role of dynamic, conversational assessment during instruction. Both studies emphasize adapting teaching to individual student needs, suggesting that integrating tools like SOLO taxonomy with real-time practices offers a comprehensive strategy for tackling diverse learning challenges, especially in complex subjects like physics.

Additionally, our findings resonate with research on textbook efficacy in teaching electrostatics(Mazibe & Rollnick, 2024). Textbooks often fail to scaffold complex concepts effectively; mirroring students' struggles to progress beyond prestructural understanding. Addressing foundational misconceptions through improved instructional materials and scaffolding can aid the transition to relational and extended abstract thinking, as outlined by the SOLO taxonomy.

Furthermore, a comparable investigation conducted in senior high schools on electrostatic concepts identified electric potential as one of the least understood topics among students, indicating a need for adjustments in teaching methods and curricula(Mi et al., 2023). Despite using a different model to assess students' understanding of electric potential, the results were similar. Another study, which analyzed questionnaires from 141 Spanish university students, found that only a small number could answer questions related to potential differences without error. This may be due to the complexity of using microscopic-level explanations involving charges or electrons (Leniz et al., 2017). Various studies have also highlighted the challenges students face in visualizing potential differences, often presenting it as an abstract concept. While this study focuses on the challenging area , other research explores solutions, such as augmented reality(Ropawandi et al., 2022) or inquiry-based learning (Kock et al., 2015) to address these challenges(Rembach & Dison, 2016). The SOLO framework, as applied in the present study, assists educators in identifying challenging area about potential differences, suggesting that this taxonomy could be a valuable tool for assessing students' depth of knowledge across various scientific topics. Further research could extend this analysis to different age groups to evaluate learning outcomes more broadly.

**Conclusion**

Potential difference is one of the most challenging concepts in physics classes, particularly due to persistent student difficulties. Addressing these challenges early is crucial. This study examines how high school textbooks in grades 11 and 12 present this topic. To evaluate student learning outcomes, the researcher employed the SOLO taxonomy, marking the first application of this model in physics education. The study revealed lack of deep learning in this area. These findings can inform teaching strategies to better support students. However, the study's small sample size limits its generalizability. Further research, particularly with larger samples and diverse educational contexts, is needed to validate the SOLO taxonomy as an effective tool in physics education

**Implication:**

Our study highlights the difficulties students face in understanding potential difference (PD), a fundamental concept in electricity. The abstract nature of PD contributes to these challenges, necessitating more effective teaching strategies (Hu, 2024). In this research, we implemented an instrument to evaluate students' depth of understanding of PD in senior high school physics. Our findings demonstrated the effectiveness of the SOLO taxonomy in assessing student learning. Additionally, we emphasized the challenges within the current syllabus. By implementing similar practices, teachers can identify and correct difficulties among students before these challenges become deeply ingrained(Tindan & Arthur, 2024).

**Limitation and further studies:**

This study was limited by the small sample size. Future research should involve a larger sample to yield more accurate results. Additionally, due to time constraints, we interviewed only two out of 92 students. Increasing the number of interviews in subsequent studies would provide a more comprehensive understanding of student challenges. Finally, our focus was on a single topic in physics textbooks; future research should explore a broader range of topics in physics.

## Acknowledgment

A special thanks goes to Professor Sanjay Rebello for their expert guidance and insightful comments, which significantly enhanced the quality of our research. We are also grateful to Masoumeh Shahsavari for her dedicated work in collecting data and offering her unique viewpoint.

We also employed AI technology to identify and correct the grammatical mistakes in the text.

## Funding details

We would like to clarify that no external funding was utilized in the preparation and publication of our papers.

## Conflict of interest

The authors report there are no competing interests to declare

## Data Availability Statement

We used available data through the local setting, which was provided by our colleagues in different classes, among 92 students in grades 11 and 12 of high school. Due to privacy concerns, the graded papers of the students, which are kept in printed form, cannot be shared publicly. However, summary statistics and findings can be accessed by contacting the author directly at [rahmanif@purdue.edu](mailto:rahmanif@purdue.edu)

## Ethic statement

This study was approved by professor Modjtaba Ghorbani, Vice President for Research and Technology in the Department of Physics ensuring that all procedures performed were in accordance with the ethical standards of the institution and the national research committee. Informed consent was obtained from all individual participants included in the study.